\documentclass[prd,floatfix,amsmath,nofootinbib,amssymb,floatfix]{revtex4-2}
\usepackage{graphicx}
\usepackage{bm}
\usepackage{amsmath}
\usepackage{amssymb}
\usepackage{hyperref}

\newcommand{\qslash}{\kern 0.2 em n\kern -0.50em /}
\newcommand{\nslash}{\kern 0.2 em n\kern -0.50em /}
\newcommand{\kslash}{\kern 0.2 em k\kern -0.45em /}
\newcommand{\lslash}{\kern 0.2 em l\kern -0.50em /}
\newcommand{\pslash}{\kern 0.2 em p\kern -0.50em /}
\newcommand{\Sslash}{\kern 0.2 em S\kern -0.50em /}
\newcommand{\Pslash}{\kern 0.2 em P\kern -0.50em /}
\newcommand{\Dslash}{\kern 0.2 em D\kern -0.65em /\kern 0.15em}

\newcommand{\eps}{\epsilon}

\newcommand{\Tr}{\operatorname*{Tr}\nolimits}

\newcommand{\ii}{i}

\usepackage{color}

\usepackage{overpic}
\usepackage{amssymb}
\usepackage{indentfirst}
\usepackage{feynmf}   
\usepackage{slashed}  
\usepackage{cases}
\usepackage{color}
\usepackage{multirow}
\usepackage{epstopdf}
\usepackage{graphicx,color,bm}
\usepackage{epstopdf}

\begin{document}
\title{The transverse-spin asymmetry $A^{\sin\phi_{S_\Lambda}}_{UUT}$ in $\Lambda$ production in SIDIS within the collinear framework}
\author{Yongliang Yang}
\email{yangyl@qdu.edu.cn (corresponding author)}
\affiliation{Centre for Theoretical and Computational Physics, College of Physics, Qingdao University, Qingdao 266071, China}
\author{Zhun Lu}
\email{zhunlu@seu.edu.cn (corresponding author)}
\affiliation{School of Physics, Southeast University, Nanjing 211189, China}
\begin{abstract}
  We investigate the transverse-spin asymmetry $A^{\sin\phi_{S_\Lambda}}_{UUT}$ for transversely polarized $\Lambda$ production in semi-inclusive deep inelastic scattering with an unpolarized electron beam and an unpolarized nucleon target.
  The spin-dependent asymmetry is sensitive to the twist-3 T-odd quark-gluon-quark fragmentation function $\tilde{D}_{T}$, which describes the correlation between an unpolarized fragmenting quark and the transverse polarization of the produced hadron.
  We calculate $\tilde{D}_{T}$ of the $\Lambda$ hyperon within a diquark model and investigate its contribution to the asymmetry.
  We determine the model parameters by fitting the fragmentation functions $D_1$ and $G_1$ to the DSV parametrization.
  Using the numerical results of the model and the available parametrization of the unpolarized parton distribution function $f_1$, we predict the asymmetry in the electroproduction of the $\Lambda$ hyperon in the kinematic regions of the EIC and the EicC.
  We also examine the impact of including the strange-quark contribution on the transverse-spin asymmetry.
  In our phenomenological analysis, we include the QCD evolution of the parton distribution function $f_1$ and the fragmentation function $\tilde{D}_{T}$.
  The results show that the asymmetry is sizable and can be significantly modified by the strange-quark contribution.
  \end{abstract}
  \maketitle

\emph{\section{introduction}}

The study of fragmentation functions (FFs), which encode the nonperturbative dynamics of hadronization, is an essential component of our understanding of high-energy particle processes, both theoretically and experimentally~\cite{Berman:1971xz,Metz:2016swz}.
FFs describe the fragmentation of a parton into an observed hadron and provide a crucial link between partonic dynamics and experimentally measurable final states.
They play an important role in the study of spin and azimuthal asymmetries in a wide range of processes, including $e^+\ e^-$ annihilation, semi-inclusive deep inelastic scattering (SIDIS), and proton-proton collisions
~\cite{Abe:2005zx,Seidl:2008xc,TheBABAR:2013yha,Ablikim:2015pta,
Guan:2018ckx,Airapetian:2004tw,Airapetian:2010ds,Qian:2011py,
Adolph:2014zba,Lesnik:1975my,Bunce:1976yb,Adams:2003fx,Abelev:2008af,
Lee:2007zzh,Adamczyk:2012xd}.

Among FFs, the unpolarized collinear FF $D_1^{h/i}(z)$ is by far the most extensively studied~\cite{Albino:2008fy,Borsa:2022vvp,Borsa:2021ran,deFlorian:2017lwf,
deFlorian:2014xna,Gao:2025bko,Gao:2025hlm}.
It describes the fragmentation of an unpolarized parton of flavor $i$ into an unpolarized hadron $h$ in the current fragmentation region.
In contrast, spin-dependent FFs, which are directly related to the polarization of the produced hadron, remain much less constrained.
Two well-known examples are the naive time-reversal-odd (T-odd) Collins function $H_1^\perp$~\cite{Wang:2018wqo,Collins:1992kk} and the Sivers-type fragmentation function $D_{1T}^\perp$~\cite{Li:2020oto,Yang:2017cwi,Anselmino:2000vs,
DAlesio:2020wjq,Callos:2020qtu}.
These functions describe correlations between the transverse spin and transverse momentum of the fragmenting quark and the produced hadron, such as a pion or a $\Lambda$ hyperon.
SIDIS therefore provides an important framework for probing both parton distribution functions (PDFs) and spin-dependent FFs, which give rise to a variety of spin and azimuthal asymmetries and provide an approach of investigating the polarization of the $\Lambda$ hyperon~\cite{Boer:1997nt,Yang:2016mxl}.
The future Electron-Ion Collider (EIC) and the proposed Electron-Ion Collider in China (EicC) will provide unprecedented opportunities to study $\Lambda$ polarization and the underlying spin-dependent fragmentation dynamics.

In recent years, twist-3 FFs arising from quark-gluon-quark (qgq) correlations have attracted considerable interest in the study of transverse single-spin asymmetries (SSAs)~\cite{Yang:2021zgy,Koike:2017fxr,Kanazawa:2014dca,Kang:2010zzb,
Kang:2010xv,Eguchi:2006qz,Qiu:1998ia}, particularly in hadron production in $pp$ collisions~\cite{Qiu:1991wg,Qiu:1998ia,Efremov:1984ip,Efremov:1981sh,Bunce:1976yb}. Despite their theoretical significance~\cite{Metz:2002iz,Collins:2004nx,Boer:2003cm,Yuan:2007nd,
Gamberg:2008yt,Meissner:2008yf}, our knowledge of twist-3 FFs remains rather limited.
In general, the extraction of FFs from experimental data is more challenging than that of PDFs. Model calculations can nevertheless provide valuable insight into their qualitative behavior and flavor dependence, particularly for higher-twist quantities such as twist-3, naive T-odd FFs. Phenomenological analyses and model studies of the $\sin\phi_S$ asymmetry in SIDIS indicate that the T-odd twist-3 FF $\tilde{H}$ should be taken into account in the description of SSAs~\cite{Lu:2015wja,Bacchetta:2006tn,Wang:2016tix,Metz:2016swz,
Yang:2021zgy}.

In addition to $\tilde H$, another twist-3 T-odd qgq FF, $\tilde D_T$, can generate a transverse-spin asymmetry involving a transversely polarized final-state hadron.
In Ref.~\cite{Yang:2016qsf}, we derived the corresponding spin-dependent asymmetry, denoted by $A^{\sin\phi_{S_\Lambda}}_{UUT}$.
This asymmetry is sensitive to the twist-3 FF $\tilde D_T$ and, within the collinear framework, can be expressed in terms of the convolution $f_1^q(x)\otimes \tilde D_T^q(z)$.
Since the dominant component of the $\Lambda$ hyperon contains  $u, d, s$ valence quarks, studying $\tilde D_T$ for $\Lambda$ production can provide complementary information on the hadronization mechanism and may offer a novel way for probing the strange-quark PDF $f_1^s$ in the proton. In particular, the asymmetry considered here involves the flavor combination
$$
\sum_q f_1^q(x)\tilde D_T^{q\to\Lambda}(z),
$$
through which the strange-quark distribution can contribute via the $s\to\Lambda$ fragmentation channel. A quantitative understanding of the flavor dependence of $\tilde D_T$ is therefore important for assessing the sensitivity of this observable to the strange-quark content of the proton.

In this work, we investigate the twist-3 T-odd FF $\tilde D_T$ for the $\Lambda$ hyperon and its contribution to the SSA $A^{\sin\phi_{S_\Lambda}}_{UUT}$ within a diquark spectator model.
The model parameters are determined by simultaneously fitting the unpolarized FF $D_1$ and the helicity FF $G_1$ to the DSV parametrization.
Based on the  model resulting FFs, we calculate the $\sin\phi_{S_\Lambda}$ asymmetry for $\Lambda$ electroproduction at the kinematics relevant to the EIC and EicC.
Since the characteristic scale of the spectator model is substantially lower than the hard scales of these experiments, we further estimate the scale dependence of $\tilde D_T$
 phenomenologically. This allows us to assess the impact of evolution effects and to facilitate a meaningful comparison of the model predictions with observables evaluated at different energy scales.

The remainder of this paper is organized as follows.
In Sec.~\ref{sec:mod}, we calculate the twist-3 qgq fragmentation function $\tilde D_T$ within the diquark spectator model and investigate its scale dependence.
In Sec.~\ref{sec:Auut}, we formulate the $A^{\sin\phi_{S_\Lambda}}_{UUT}$ asymmetry in SIDIS with the transverse momentum of the final-state hadron integrated over.
We then present numerical predictions for the $\sin\phi_{S_\Lambda}$ asymmetry in $\Lambda$ electroproduction at the EIC and EicC within the collinear framework.
Finally, in Sec.~\ref{sec:Con}, we summarize our main results and draw our conclusions.

\emph{\section{The collinear fragmentation function $\tilde{D}_{T}$ in a diquark model}\label{sec:mod}}

\subsection{Model calculation of $\tilde{D}_{T}$ }

\begin{figure}
  \centering
  \includegraphics[width=7.5cm]{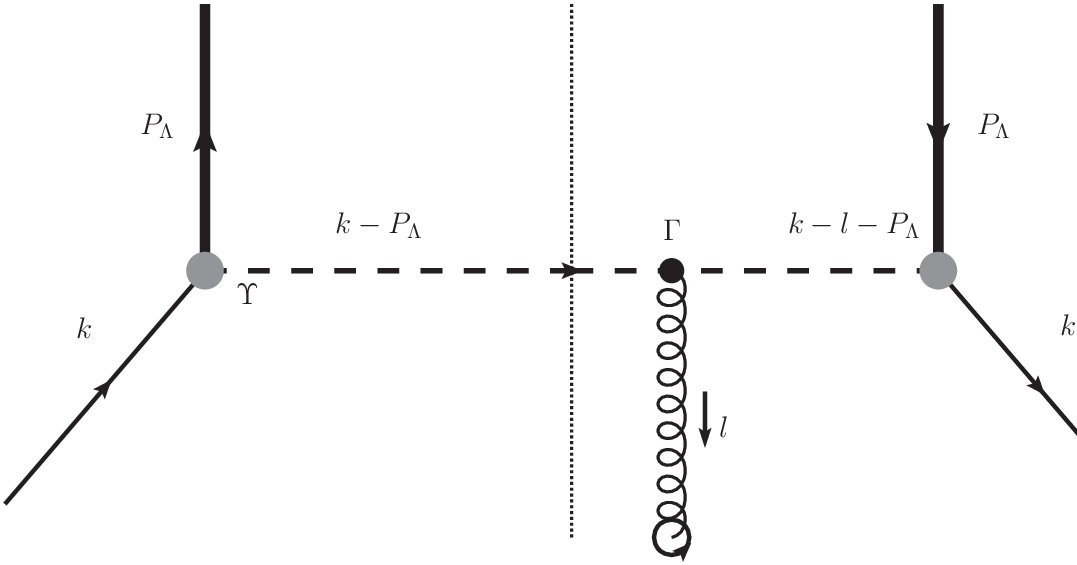}
  \caption{The Feynman diagram is relevant to the calculation of the quark-gluon-quark correlator in the diquark model.}\label{qgqmc}
\end{figure}

In this section, we calculate the twist-3 transverse-momentum-dependent (TMD) FF $\tilde{D}_{T}(z,\bm{k}^2_T)$ in a diquark spectator model.
The fragmentation function can be extracted from the following trace of the quark-gluon-quark correlator~\cite{Yang:2017cwi,Bacchetta:2006tn}:
\begin{align}
{z\over 2M_\Lambda}\Tr[\tilde{\Delta}_{A\alpha}(g_T^{a\alpha}
-\ii\eps_T^{a\alpha}\gamma_5)\gamma^-]&=\eps^{aS_T}(\tilde{D^\prime}_T
+\ii\tilde{G^\prime}_T)
-{k_T\cdot S_T\over M_\Lambda^2}\eps^{ak_T}(\tilde{D^\perp_T}+\ii\tilde{G^\perp_T})\,,
\label{eq:trace}
\end{align}
where the twist-3 FF entering the observable is obtained from the combination
\begin{align}
\tilde{D}_T=\tilde{D^\prime}_T-{k^2_T\over 2M^2_\Lambda} \tilde{D}^\perp_T\,, \label{eq:DT}
\end{align}
and the twist-3 spin-dependent qgq fragmentation correlator $\tilde{\Delta}_{A\alpha}$ can be expressed as~\cite{Gamberg:2006ru,Bacchetta:2006tn}
\begin{align}
\tilde{\Delta}_A^\alpha(z,k_T;S_{\Lambda}) &=\sum_{X}\hspace{-0.55cm}\int \; \frac{1} {2z}\int \frac{d\xi^{+}d^2\bm\xi_T} {(2\pi)^3}\int  e^{\ii k\cdot \xi} \langle 0| \int^{\xi^+}_{\pm\infty^+} d{\eta^+}\mathcal{U}^{\bm\xi_T}_{(\infty^+,\eta^+)}\nonumber\\
 &\times gF^{-\alpha}_\perp (\eta) \mathcal{U}^{\bm\xi_T}_{(\eta^+,\xi^+)} \psi(\xi)|P_{\Lambda},S_{\Lambda};X\rangle\langle P_{\Lambda},S_{\Lambda};X|\bar{\psi}(0)\mathcal{U}^{\bm 0_T}_{(0^+,\infty^+)}\mathcal{U}^{\infty^+}_{(\bm 0_T,\bm \xi_T)}|0\rangle\bigg|_{\begin{subarray}{l}
\eta^- = \xi^-=0 \\ \eta_T = \xi_T \end{subarray}}\,.
\label{eq:qgq}
\end{align}
where the light-cone coordinates $a^{\pm}=a\cdot n_{\pm}=(a^0\pm a^3)/\sqrt{2}$ have been applied, and $k^{-}={P_\Lambda^{-}/{z}}$.
$k$ and $P_\Lambda$ denote the momenta of the parent quark and produced hadron, respectively.
The spin-dependent fragmentation function $\tilde{D}_T$ can be calculated from Fig.~\ref{qgqmc}.
In the diquark model~\cite{Nzar:1995wb,Jakob:1997wg}, the correlator is expressed as~~\cite{Yang:2021zgy,Lu:2015wja}:
\begin{align}\label{eq:qgqFD}
\tilde{\Delta}^\alpha_A={C_F\alpha_S\over 4(2\pi)^22(1-z)P_\Lambda^-}{1\over k^2-m^2}\int{d^4 l\over (2\pi)^4}{(l^-g_T^{\alpha\rho}-l^\alpha\,n_+^{\rho})(\kslash - \lslash + m)\bar{\mathcal{Y}}^\nu(\Pslash_\Lambda + M_\Lambda)(\gamma_5\Sslash_\Lambda)
\mathcal{Y}^\mu(\kslash+m)\over((k-l)^2-m^2)(l^2-\ii\eps)
((k-l-P_\Lambda)^2-m_s^2)(-l^-\pm\ii\eps)} d^{i\mu} d^{j\nu} \bar{\Gamma}_{i\,j\rho},
\end{align}
where the notation $\mathcal{Y}$ describes the scalar and axial vector form of quark-diquark-hyperon and can be expressed as
${\mathcal Y }_s=g_s(k^2)$
and ${\mathcal Y }^{\mu}_v ={g_v(k^2)\over\sqrt{3}}\gamma_5(\gamma^\mu+{P_\Lambda^\mu\over M_\Lambda})$, respectively.
The polarization sum for the axial-vector diquark is chosen as $d^{\mu\nu}=\sum_{\lambda_a} \epsilon^\mu_{\lambda_a}\epsilon^\nu_{\lambda_a}=-g^{\mu\nu}
+{P_\Lambda^\mu\,P_\Lambda^\nu\over M_\Lambda^2}$, and the gluon-diquark coupling vertex $\Gamma$ with scalar and axial vector diquark:
\begin{align}
\Gamma^{\rho} = &\ii\,(2k-2P_\Lambda-l)^\rho;\\
\Gamma^{\rho\mu\nu} = &-\ii\,[(2k-2P_\Lambda-l)^\rho\,g^{\mu\nu} -(k-P_\Lambda-l)^\mu g^{\nu\rho}-(k-P_\Lambda)^\nu g^{\rho\mu}].
\end{align}
where the anomalous chromomagnetic moment is set to $\kappa=0$ in the axial-vector-diquark vertex.
The Feynman rule corresponding to the gluon field strength tensor $F^{\alpha\beta}$ is given by the factor $\ii\,(l^\alpha g_T^{\rho\beta}-l_T^\alpha g^{\beta\rho})$ as denoted by the open circle in Fig.~\ref{qgqmc}~\cite{Lu:2015wja}.

To obtain the imaginary part of the correlator, we utilize the Cutkosky cut rule to put the gluon and quark lines on the mass shell.
A similar calculation for the T-odd qgq FF $\tilde{H}$ can be found in Ref.~\cite{Yang:2021zgy}.
This corresponds to the following replacements on the propagators by using the Dirac delta functions
\begin{align}
{1\over l^2 + \ii\varepsilon} \rightarrow -2\pi i\delta(l^2),~~~~~~~ {1\over (k-l)^2-m^2+\ii\varepsilon} \rightarrow -2\pi i\delta((k-l)^2-m^2) \,.\label{eq:cuts}
\end{align}

Using Eqs.~\eqref{eq:trace}, \eqref{eq:DT}, and \eqref{eq:qgqFD}, we obtain the scalar- and axial-vector-diquark contributions to $\tilde D_T(z,\bm k_T^2)$ for the $\Lambda$ hyperon as follows:
\begin{align}
\tilde{D}^S_T(z,\bm{k}_T^2)={C_F\alpha_S\over 4\,M_\Lambda(2\pi)^2(1-z)}{1\over k^2-m^2}\tilde{D}_{TS}(z,\bm{k}_T^2),
\end{align}
where the trace result of integral is given by:
\begin{align}
 \tilde{D}_{TS}(z,\bm k_T^2)&=2 g_s^2(z\,m+M_\Lambda) \{\mathcal{A}[M_\Lambda^2(z-1)-z m_D^2]+\mathcal{B} M_\Lambda^2 (z-1)-\bm{k}^2_T z[\mathcal{C} k^-(1-z) +\mathcal{I}_2]\}.
\end{align}

Similarly, the expression of the axial-vector parts reads:
\begin{align}\label{eq:MRV}
\tilde{D}^V_T(z,\bm{k}_T^2)={z\,C_F\alpha_S\over 4\,M_\Lambda(2\pi)^2(1-z)}{1\over k^2-m^2}\{\tilde{D}_{TV}(z,\bm{k}_T^2)+\tilde{D}_{TV0}(z,\bm{k}_T^2)
+\tilde{D}_{TV1}(z,\bm{k}_T^2)
+\tilde{D}_{TV2}(z,\bm{k}_T^2)\},
\end{align}
where the four terms in the r.h.s. of Eq.(\ref{eq:MRV}) are given by
\begin{align}
 \tilde{D}_{TV}& =-{2g_v^2\over 3z}(z\,m+M_\Lambda) \{\mathcal{A}[M_\Lambda^2(z-1)-z m_D^2]+\mathcal{B} M_\Lambda^2 (z-1)-\bm{k}^2_T z[\mathcal{C} k^-(1-z) +\mathcal{I}_2]\},\notag\\
 \tilde{D}_{TV0}&={2g_v^2\over 3 M_\Lambda} \{(k\cdot P_\Lambda+m M_\Lambda) [(\mathcal{AA}+z\mathcal{AB}) k\cdot P_\Lambda+M_\Lambda^2 (\mathcal{AB}+z\mathcal{BB} )+z\mathcal{W}_1]+\bm{k}^2_T M_\Lambda^2(\mathcal{AA}+z\mathcal{AB})\notag\\
 &-(\mathcal{A}+z\mathcal{B})[2m M_\Lambda k\cdot P_\Lambda+(k\cdot P_\Lambda)^2+M_\Lambda^2 (m^2+2\bm{k}^2_T)]\},\notag\\
 \tilde{D}_{TV1}&={g_v^2\over 3 M_\Lambda} \bigg{\{}M_\Lambda \{z [-\bm{k}^2_T(m\mathcal{AA} +2M_\Lambda\mathcal{AB}-4m\mathcal{A}-2M_\Lambda\mathcal{B})+(M_\Lambda
 \mathcal{B}-m\mathcal{I}_2)(k^2+3 m^2)+2m\mathcal{W}_1]\notag\\
 &+2 M_\Lambda[-\mathcal{AA}\bm{k}^2_T+3 \mathcal{A}\bm{k}_T^2+M_\Lambda (M_\Lambda\mathcal{BB}-m\mathcal{B})+3\mathcal{W}_1]\}\notag\\
 &+k\cdot P_\Lambda \{z[-\mathcal{AA}\bm{k}^2_T+(\mathcal{A}-\mathcal{I}_2) (k^2+3 m^2)-2\mathcal{W}_1]+2k\cdot P_\Lambda(\mathcal{AA}-3\mathcal{A}+2 \mathcal{I}_2)\notag\\
 &+2 M_\Lambda (2\mathcal{AB}M_\Lambda-\mathcal{A}m-3\mathcal{B} M_\Lambda+2\mathcal{I}_2 m)\}\bigg{\}},\notag\\
 \tilde{D}_{TV2}&=-{g_v^2\over 3 M_\Lambda^2 }\bigg{\{}{z^2\over 2} \{-\bm{k}^2_T[\mathcal{CC}k^- (k^2-m^2)(m+M_\Lambda)+2 m M_\Lambda (\mathcal{CD}k^- M_\Lambda-2\mathcal{C}k^- m)-m{(k^2-m^2)\over 2k^2}\mathcal{I}_1]\notag\\
 &+\mathcal{W}_2k^-(k^2-m^2)(m+M_\Lambda)+m{k^2-m^2\over 2}\mathcal{I}_1\}\notag\\
 &+M_\Lambda^2 z[-\bm{k}^2_T(2\mathcal{CC}k^-(z m+M_\Lambda)+\mathcal{CD}k^- M_\Lambda z)+4\mathcal{W}_2k^-(z m+M_\Lambda)]\notag\\
 &+z \{-\bm{k}^2_T \mathcal{CE}k^-k^- z (M_\Lambda-z m)-k\cdot P_\Lambda [4\mathcal{W}_2k^-(M_\Lambda+z m)+\bm{k}^2_T(2\mathcal{C}k^- z m-\mathcal{CC}k^- (3z m+M_\Lambda))]\}\bigg{\}}\notag.
\end{align}
Here, the subscript $i$ of $\tilde D_{TV,i}$ denotes the power of the factor $l^-$ appearing in the numerator after the trace in Eq.~\eqref{eq:MRV} is evaluated.
These functions $\mathcal{A}$, $\mathcal{B}$, $\mathcal{C}$, $\mathcal{I}_i$, $\mathcal{AA}$, $\mathcal{AB}$, $\mathcal{BB}$, $\mathcal{CC}$, $\mathcal{CD}$, $\mathcal{CE}$ and $\mathcal{W}_i$ depend on $k^2$, $m$, $m_D$, and $M_\Lambda$.
Explicit expressions for some of these functions can be found in Refs.~\cite{Yang:2021zgy,Yang:2017cwi}.
In particular, $\mathcal{W}_2k^-$ is given by
\begin{align}
\mathcal{W}_2k^-={2 (\mathcal{A} k^2+\mathcal{B} k\cdot P_\Lambda)-\mathcal{C}k^- (z-1) (k^2-m^2)\over 2 z }.
\end{align}

For the scalar and axial vector hyperon-quark-diquark coupling factors $g_s$ and $g_v$, we take $g_s$ and $g_v$ to have the same $k^2$-dependent Gaussian form, namely $g_s=g_v=g_D(k^2)$, with
 \begin{align}\label{factor}
  g_D(k^2)= {g_{qh}\over z}\,e^{-{k^2\over \Lambda^2}}\,,
 \end{align}
where $\Lambda^2$ has the general form $\Lambda^2=\lambda^2z^\alpha(1-z)^\beta$.
The Gaussian form factor is introduced to regularize the divergence in the quark momentum integration.
The choice in Eq.~\ref{factor} has the additional benefit of providing a reasonable description of the unpolarized FF $D_1$.
Assuming SU(6) spin-flavor symmetry, the FFs of the $\Lambda$ hyperon for light flavors satisfy the following relations~\cite{VanRoyen:1967nq,Jakob:1993th}
 \begin{align}\label{relation}
 D^{u\rightarrow \Lambda} =\,D^{d\rightarrow \Lambda} ={1\over 4}D^{(S)}+{3\over 4}D^{(V)}\,,~~D^{s\rightarrow \Lambda}=D^{(S)}\,,
 \end{align}
where $u, d$ and $s$ denote the up, down and strange quark flavors, respectively.
Symmetry-breaking effects in polarization-related $\Lambda$ hyperon production are considered to be small~\cite{Chen:2021hdn}.
In this study, we therefore assume that the relation in Eq.~(\ref{relation}) holds for all FFs.
This assumption is also compatible with the DSV parametrization for $D_1$, in which all light valence quarks fragment equally into the $\Lambda$ hyperon.

\subsection{Fitting procedure for the model parameters}

The model involves the parameters $g_{qh}$, $\lambda$, $\alpha$, and $\beta$, as well as the constituent quark mass $m$ and the diquark spectator mass $m_D$.
In this paper, we fix the constituent quark mass to $m=0.36\,\mathrm{GeV}$ for the up, down, and strange quarks, and take the $\Lambda$ hyperon mass to be $1.116\,\mathrm{GeV}$.
The remaining parameters are determined through a simultaneous fit of the model results to the leading-order DSV parametrization of the unpolarized FF $D_1$ and the longitudinally polarized FF $G_1$ at the initial scale $\mu^2_{\mathrm{LO}} = 0.23\,\mathrm{GeV}^2$~\cite{deFlorian:1997zj}.
For the function $G_1(z)$, we adopt the scenario~1 parametrization, in which the up and down quark contributions vanish and only the strange-quark FF is retained.
This choice is in agreement with the results of our model calculations presented in Ref.~\cite{Yang:2017cwi}.
The fitted parameter values are summarized in Table~\ref{table1}.
  \begin{table}
\begin{tabular}{c|c|c|c|c|c}
  \hline
  ~$g_{qh}~$ & $~\lambda$ (GeV)~ &~$m_D$~ (GeV) &~ $m$ (GeV) &~ $\alpha$ & ~$\beta$  \\
\hline\hline
  $2.83^{+0.12}_{-0.1}$ & $3.567^{+0.23}_{-0.21}$ & $0.397^{+0.03}_{-0.03}$& ~0.36 (fixed)~ &~ $0.2^{+0.04}_{-0.04}$ ~ &~ $0.52^{+0.08}_{-0.08}$~ \\
  \hline
\end{tabular}
\caption{Fitted values of the parameters in the spectator model.
The value of quark mass is a fixed parameter in the fit.}\label{table1}
\end{table}

\begin{figure}
  \centering
  \includegraphics[width=0.45\columnwidth]{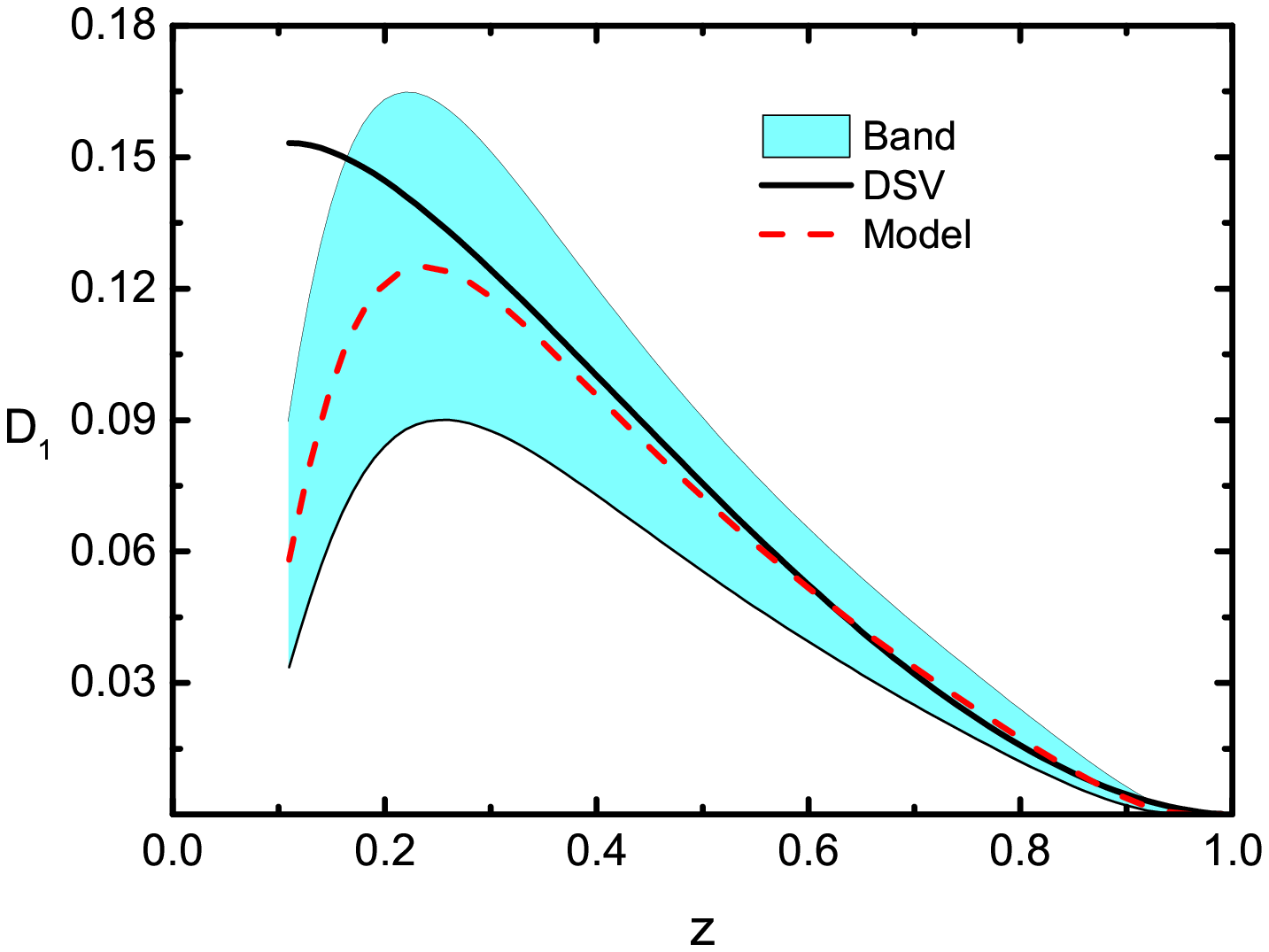}
  \includegraphics[width=0.45\columnwidth]{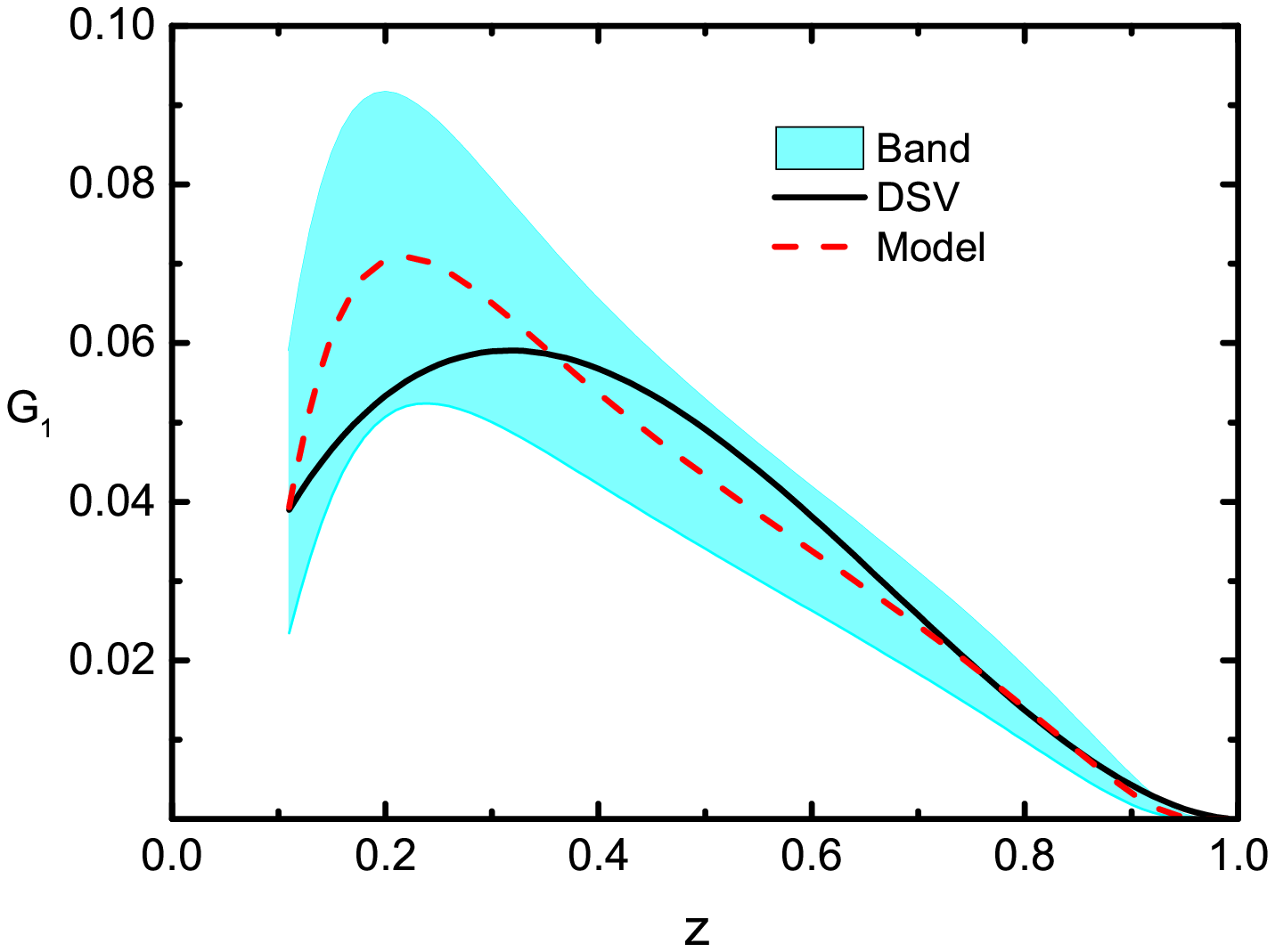}
  \caption{The fitting results of $D^{q\to\Lambda}_1(z)$ vs $z$ (left panel), the polarized fragmentation functions $G^{s\to\Lambda}_1(z)$ vs. $z$( right panel) within scenario 1. The shaded areas correspond to the uncertainty of the model parameters $g_{qh},~m_D,~\lambda$.}\label{udsfig1}
\end{figure}

Using the fitted model parameters, we present our model calculations of the FFs $D^{q\to\Lambda}_1(z)$ and $G^{s\to\Lambda}_1(z)$ as functions of $z$, shown as solid lines in the left and right panels of Fig.~\ref{udsfig1}, respectively.
For comparison, the results of the DSV parametrization~\cite{deFlorian:1997zj} are also shown as dashed curves.
The shaded bands indicate the uncertainties arising from the model parameters.
In the region $z > 0.2$, our model results are in good agreement with the DSV parametrization.

In this subsection, we present our model calculation of the collinear twist-3 FF $\tilde{D}_T(z)$ for the $\Lambda$ hyperon by integrating over the transverse momentum:
\begin{align}
\tilde{D}_T(z)=\int d^2{\bm{P}_{\Lambda\,T}}\tilde{D}_T(z,\bm{k}_T^2)=z^2\int d^2{\bm{k_T}}\tilde{D}_T(z,\bm{k}_T^2).
\end{align}
The prefactor $z^2$ appears because the transverse momentum of the produced hadron relative to the fragmenting quark is given by $P_{\Lambda T}=-zk_T$~\cite{Bacchetta:2006tn}.
In our model, we include the favored fragmentation channels ${u,\,d,\,s}\to\Lambda$, while the unfavored quark contributions are neglected.
Since experimental scales are much higher than the model scale, it is necessary to incorporate QCD evolution effects for the fragmentation functions.
The evolution of several twist-3 fragmentation functions has been studied in the literature~\cite{Kang:2010xv,Kang:2010zzb,Kang:2014zza,Ma:2017upj}.
Since the exact DGLAP evolution kernel for $\tilde{D}_T$ is not available, the present evolution effects should be regarded as a phenomenological approximation.
In this work, we assume that $\tilde{D}_T$ evolves in the same way as the unpolarized FF $D_1$, and we implement this evolution using the QCDNUM package~\cite{Botje:2010ay}.

\begin{figure}
 \centering
\includegraphics[width=0.45\columnwidth]{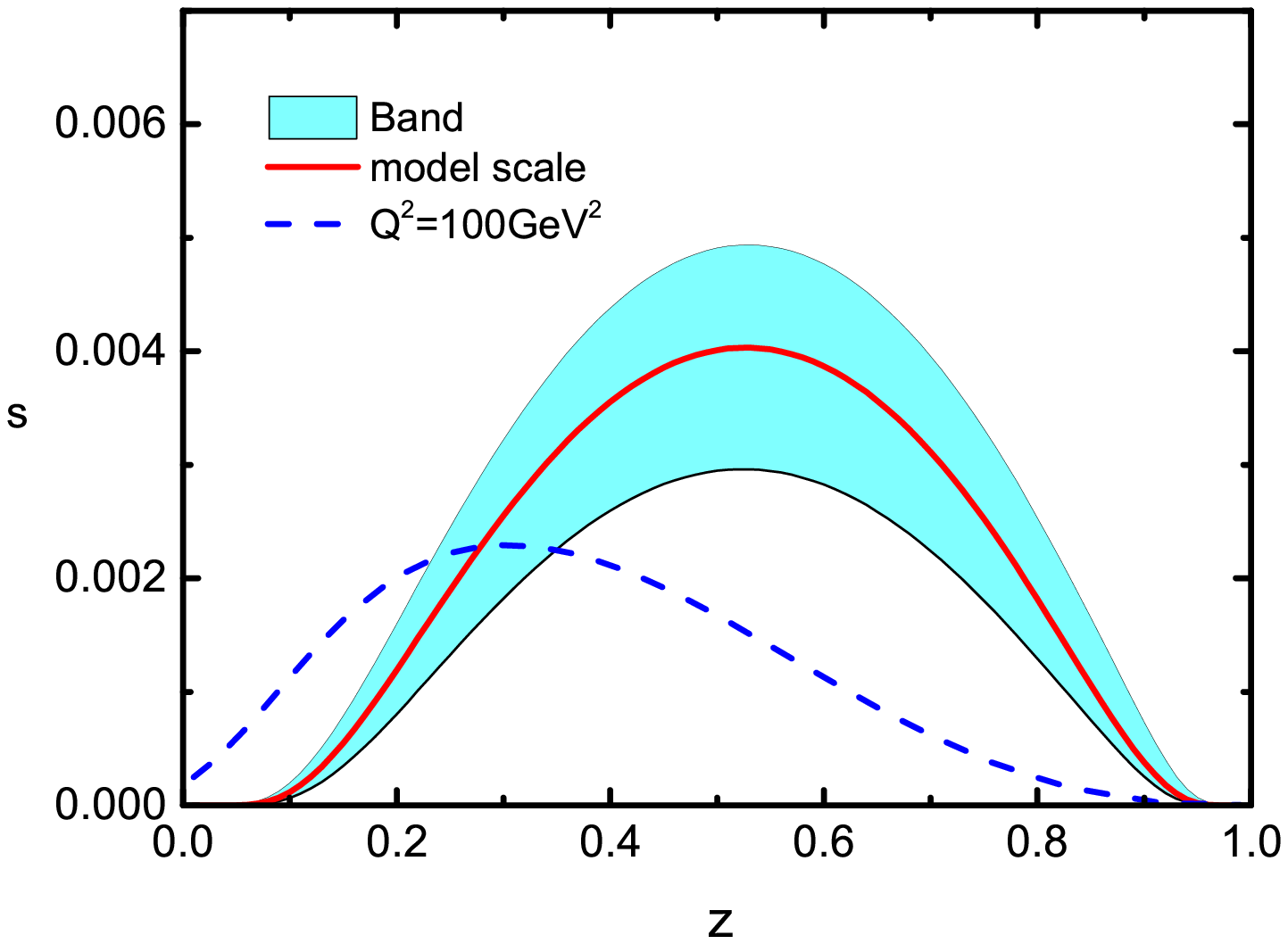}
\includegraphics[width=0.45\columnwidth]{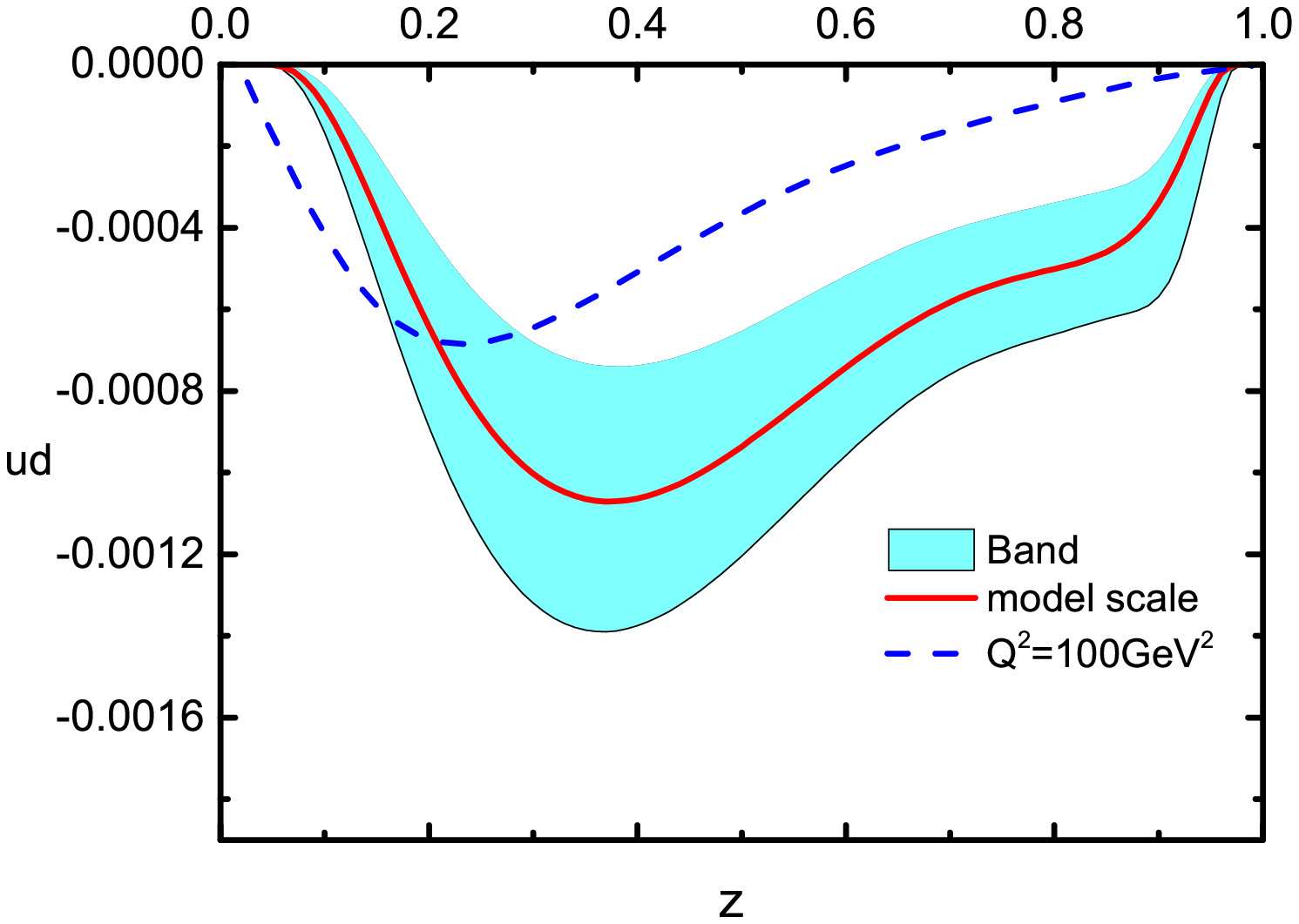}
\caption{Result of $z\tilde D_T^{\Lambda/s}(z)$ (left panel) and  $z\tilde D_T^{\Lambda/u(d)}(z)$ (right panel) at model scale $Q^2\ =0.23~\mathrm{GeV}^2$ (red solid lines) and the evolved results at $Q^2=100\,\mathrm{GeV}^2$ (blue dashed lines).}
\label{fig:Dtilde}
\end{figure}

In Fig.~\ref{fig:Dtilde}, we present the $z$-dependence and the QCD evolution effects for the collinear twist-3 FF $z \tilde{D}_T(z)$.
The left and right panels correspond to the $s$ and $u(d)$ quark channels, respectively.
The red solid and blue dashed curves denote the results at $Q^2=0.23\,\mathrm{GeV}^2$ and $100\,\mathrm{GeV}^2$.
As shown in Fig.~\ref{fig:Dtilde}, $z\tilde{D}_T(z)$ is negative for the $u(d)\to\Lambda$ fragmentation channel, while it is positive for the $s$-quark channel.
After QCD evolution from $Q^2=0.23\mathrm{GeV}^2$ to $Q^2=100 \mathrm{GeV}^2$, the magnitudes of $z\tilde D_T(z)$ are enhanced in the small-$z$ region, and the distributions are shifted toward lower $z$.
Moreover, the magnitude of the strange-quark contribution is significantly larger than that of the $u$ and $d$ quarks over a wide range of $z$, indicating that the $s\to\Lambda$ fragmentation plays a dominant role in generating this twist-3 FF.
The $z$-dependence of $z \tilde{D}_T(z)$ for $u(d)$ and $s$ quarks exhibits distinct behaviors over the full $z$ range.
This difference can be traced back to the flavor structure of the spectator model: according to Eq.~\eqref{relation}, the $s\to\Lambda$ contribution is determined by the scalar-diquark component, whereas the $u(d)\to\Lambda$ contribution receive both scalar- and axial-vector-diquark contributions.

\section{Numerical results for the transverse-spin asymmetry $A^{\sin\phi_{S_\Lambda}}_{UUT}$}\label{sec:Auut}

\begin{figure}
  \centering
  \includegraphics[width=8.5cm]{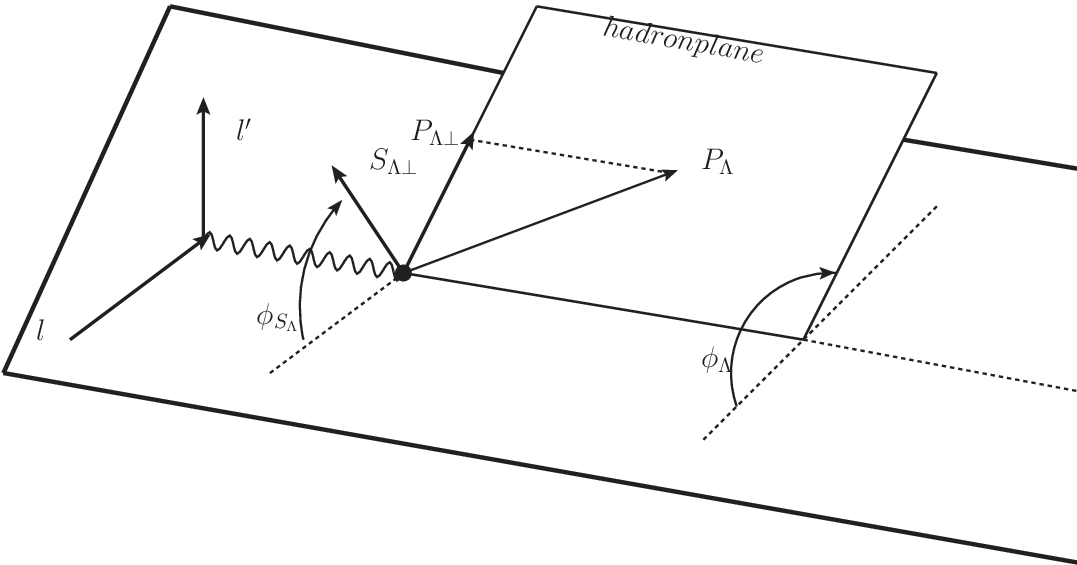}
  \caption{The definition of the azimuthal angles for SIDIS in the $\gamma^* N$ collinear frame~\cite{Yang:2016qsf}.\label{sidisplane}}
\end{figure}

In this work, we consider the production of a transversely polarized $\Lambda$ hyperon in the SIDIS process,
\begin{equation}
\label{eq:sidis}
\ell+N \longrightarrow \ell^\prime+{\Lambda}^\uparrow+X,
\end{equation}
where both the incident lepton beam and the nucleon target are unpolarized.
The reference frame adopted in this work is defined in Fig.~\ref{sidisplane}~\cite{Yang:2016qsf}.
The lepton scattering plane is spanned by the momenta $l$ and $l^\prime$~\cite{Bacchetta:2006tn}.
The transverse components of the $\Lambda$-hyperon momentum and spin vector, with respect to the virtual-photon momentum, are denoted by $P_{\Lambda\perp}$ and $S_{\Lambda\perp}$, respectively.
The azimuthal angle $\phi_{S_\Lambda}$ is defined as the azimuthal angle of the transverse spin vector $S_{\Lambda\perp}$, measured with respect to the lepton scattering plane.

Using the structure functions entering the differential cross section of the process, the $z$ dependent asymmetry can be defined as ~\cite{Wang:2016tix,Bacchetta:2006tn}
\begin{align}
A^{\sin\phi_{S_\Lambda}}_{UUT}(z)={\int dx\int dy {\alpha_{em}^2\over xyQ^2}{y^2\over 2(1-\varepsilon)}(1+{\gamma^2\over 2x})\sqrt{2\varepsilon(1+\varepsilon)} F^{\sin\phi_{S_\Lambda}}_{UUT}(x,z)\over \int dx\int dy {\alpha_{em}^2\over xyQ^2}{y^2\over 2(1-\varepsilon)}(1+{\gamma^2\over 2x})F_{UUU}(x,z)}.
\end{align}
The variables $x,y,z,\gamma$ are introduced to express the differential cross section~\cite{Bacchetta:2006tn}, where $\alpha_{em}$ is the fine structure constant and $\varepsilon$ is the ratio of the longitudinal and transverse photon flux:
\begin{equation}
\label{eq:epsilon}
\varepsilon=\frac{1-y-\frac{1}{4}\gamma^2y^2}
{1-y+\frac{1}{2}y^2+\frac{1}{4}\gamma^2y^2}.
\end{equation}
In a similar way, $x$ dependent $\sin \phi_{S_\Lambda}$ asymmetry can be written as
\begin{align}
\label{eq:autx}
A_{UUT}^{\sin\phi_{S_\Lambda}}(x)
=\frac{\int dy\int dz
\frac{\alpha_{em}^2}{xyQ^2}\frac{y^2}{2(1-\varepsilon)}(1+\frac{\gamma^2}{2x})
\sqrt{2\varepsilon(1+\varepsilon)}F^{\sin\phi_{S_\Lambda}}_{UUT}(x,z)}
{\int dy\int dz
\frac{\alpha_{em}^2}{xyQ^2}\frac{y^2}{2(1-\varepsilon)}(1+\frac{\gamma^2}{2x})
F_{UUU}(x,z)}\,,
\end{align}
where the nonvanishing integrated structure functions are given by~\cite{Yang:2016qsf,Boer:1997nt}
\begin{align}
F_{UUU}(x,z)&=x\sum_q e_q^2f_1^q(x)D_1^q(z)\,,\label{eq:FS1}\\
F^{\sin\phi_{S_\Lambda}}_{UUT}\left(x,z\right) & =  x\sum_q e_q^2\frac{2M_\Lambda}{Q}f_1^q(x)\frac{\tilde D_T^q(z)}{z}\,,\label{eq:FS2}
\end{align}
To estimate the asymmetries, we adopt the CT10 parametrization of the unpolarized PDF $f_1^q(x)$ for different quark flavors from Ref.~\cite{Lai:2010vv}.
In addition to the dominant $u$ and $d$ quark distributions, this parametrization also contains information on the strange-quark distribution $f_1^s(x)$.
Therefore, it allows us to include the strange-quark contribution consistently in the flavor summation of the structure functions.

For the EIC, we adopt the following kinematical cut~\cite{Accardi:2012qut}:
\begin{align}
& 0.001<x<0.4,\quad 0.01<y<0.95,\quad 0.2<z<0.8,\nonumber\\
& Q^2>1 \mathrm{GeV}^2, \quad \sqrt{s}=45\ \mathrm{GeV},\quad W>5\ \mathrm{GeV}.
\end{align}
Here, $W^2=(P+q)^2\approx (1-x)Q^2/x$ denotes the squared invariant mass of the virtual photon-nucleon system, while $s=(P+l)^2$ is the squared center-of-mass energy of the lepton-nucleon system.
For the EicC, we impose the following kinematic cuts~\cite{Anderle:2021wcy}:
\begin{align}
&0.005<x<0.5,\quad 0.07<y<0.9, \quad 0.2<z<0.7,\nonumber\\
&Q^2>1 \mathrm{GeV}^2,\quad \sqrt{s}=16.7\ \mathrm{GeV},\quad W>2 \mathrm{GeV}.
\end{align}
Since the EIC and EicC cover broad ranges of $Q^2$, we take into account the QCD evolution of both the PDFs and FFs in our numerical calculations.

\begin{figure}
  \centering
  \includegraphics[width=0.45\columnwidth]{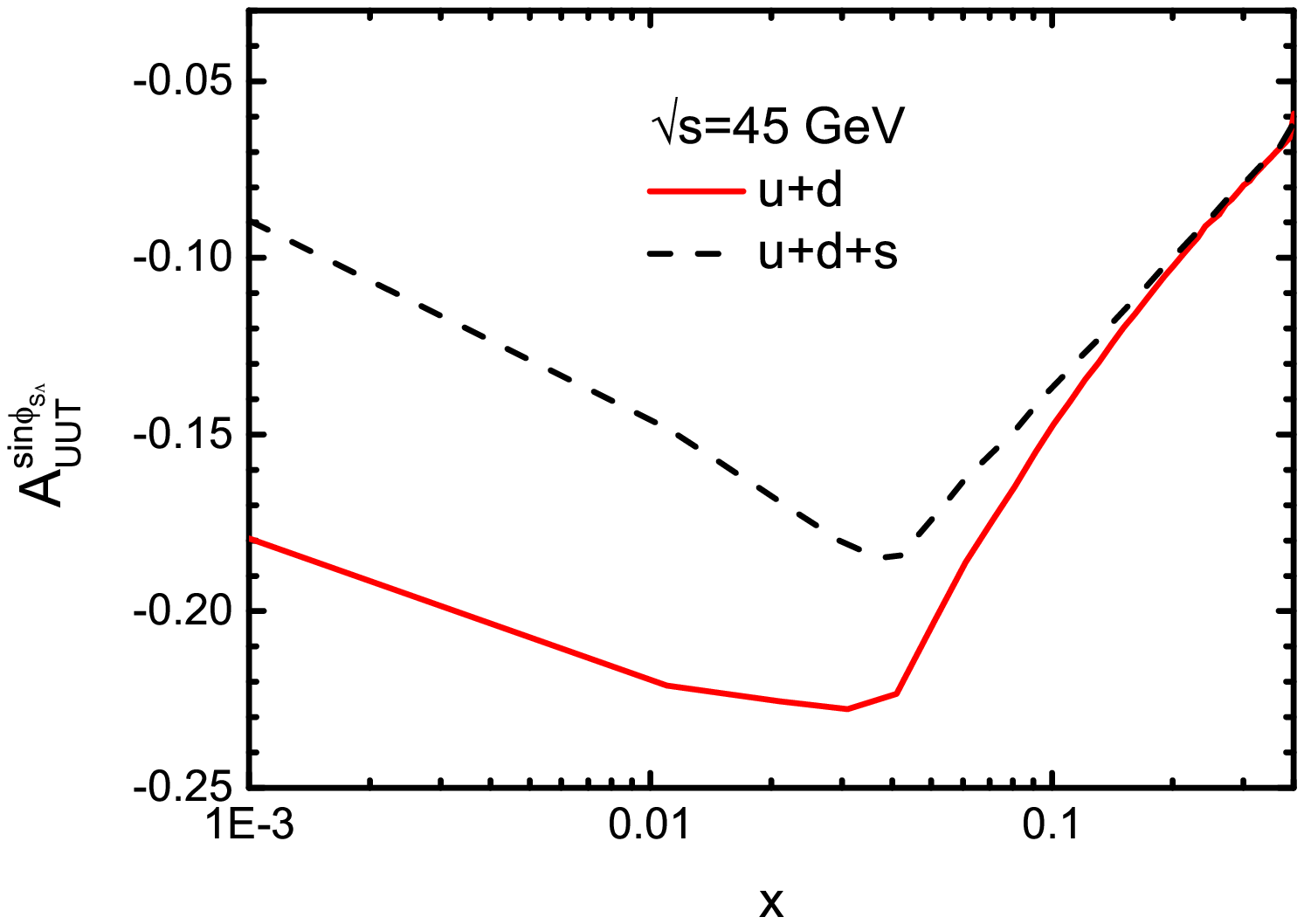}
  \includegraphics[width=0.45\columnwidth]{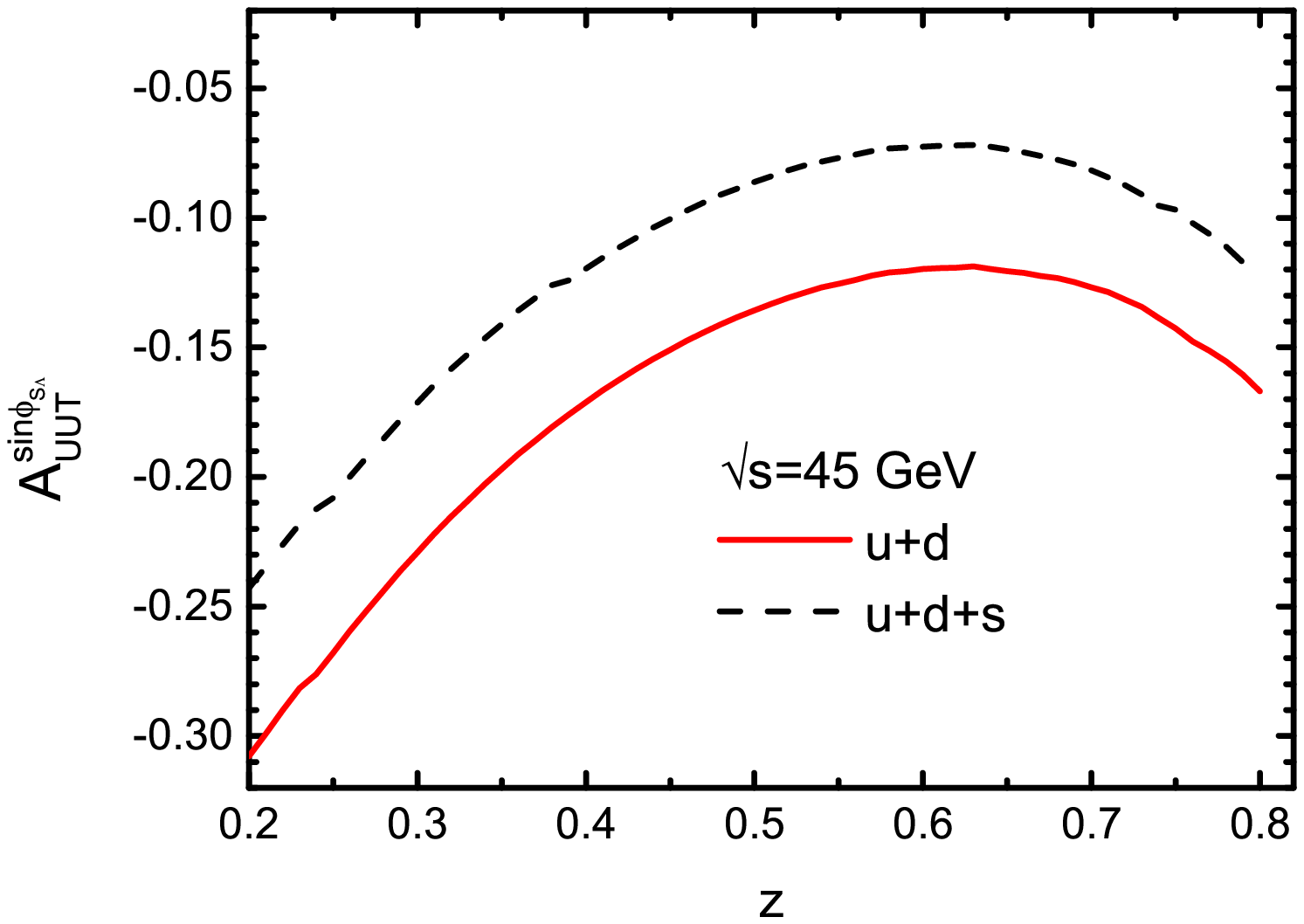}
  \caption{Transverse SSA $A^{\sin\phi_{S_\Lambda}}_{UUT}$ of $\Lambda$ hyperon production in SIDIS at EIC for $\sqrt{s}=45$ GeV. The left and the right panels show the $x$-dependent and the $z$-dependent asymmetry, respectively.}
\label{fig:eic}
\end{figure}

\begin{figure}
  \centering
  \includegraphics[width=0.45\columnwidth]{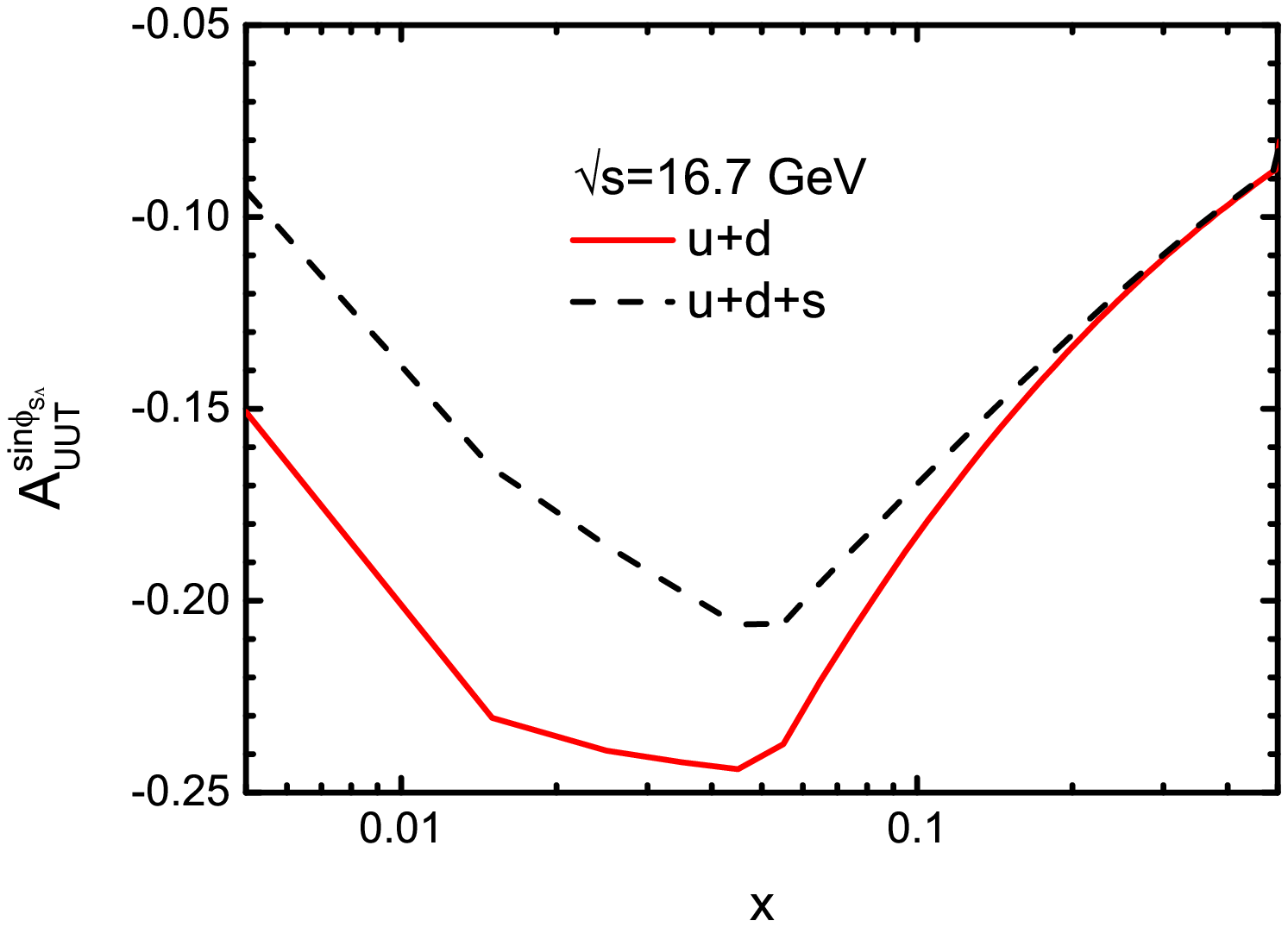}
  \includegraphics[width=0.45\columnwidth]{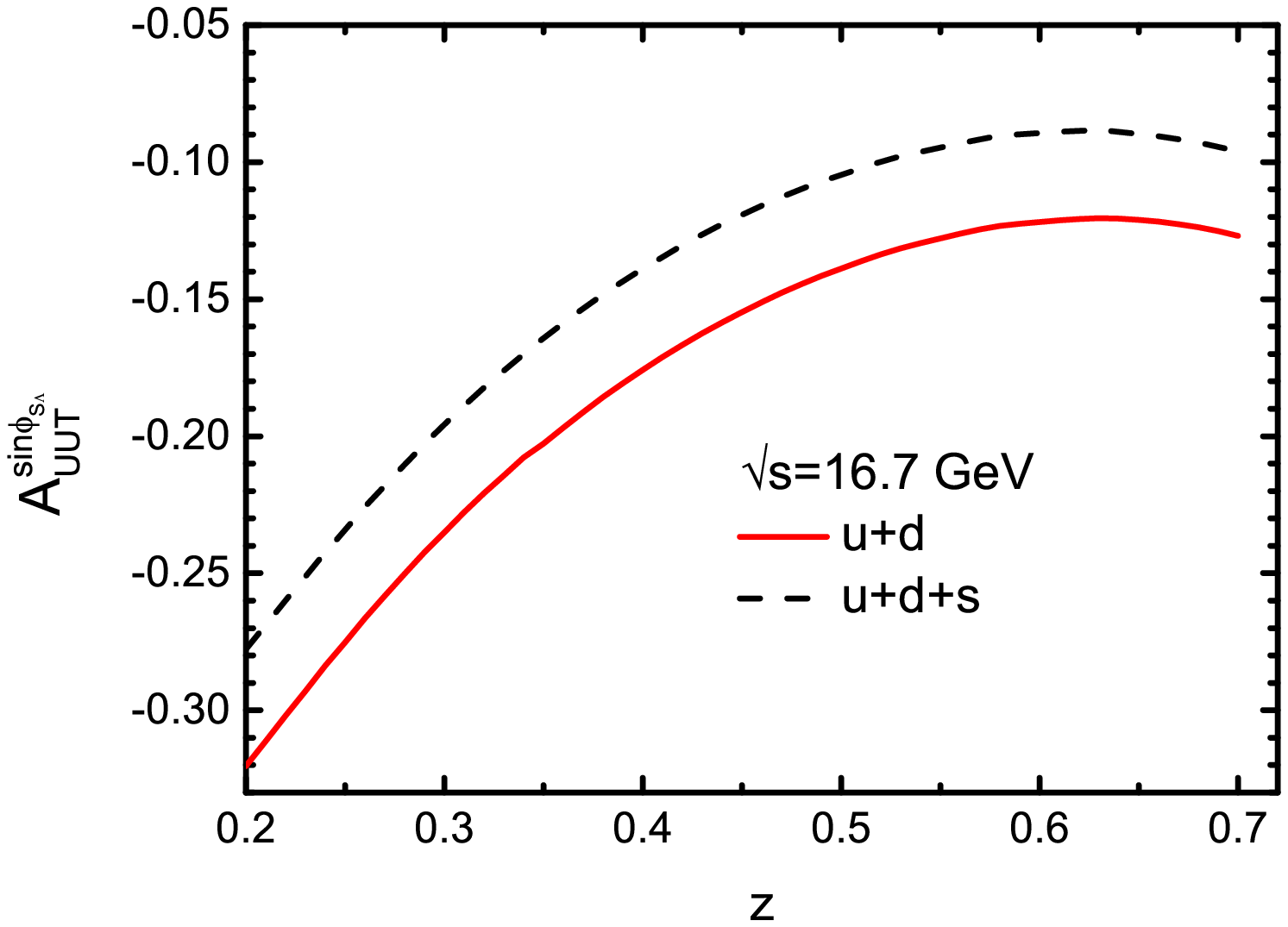}
  \caption{Transverse SSA $A^{\sin\phi_{S_\Lambda}}_{UUT}$ of $\Lambda$ hyperon production in SIDIS at EicC for $\sqrt{s}=16.7$ GeV. The left and the right panels show the $x$-dependent and the $z$-dependent asymmetry, respectively.}
\label{fig:eicC}
\end{figure}

The numerical results for the $\sin\phi_{S_\Lambda}$ asymmetries in $\Lambda$ hyperon production at EIC and EicC are shown in Figs.~\ref{fig:eic} and~\ref{fig:eicC}, respectively.
In each figure, the left and right panels display the $x$- and $z$-dependent asymmetries.
The solid curves correspond to the results including only the $u$- and $d$-quark contributions, while the dashed curves additionally include the strange-quark contribution.

We find that the predicted $\sin\phi_{S_\Lambda}$ asymmetry is negative in the considered kinematical regions and can reach a sizable magnitude.
This indicates that the transverse polarization of the produced $\Lambda$ hyperon may generate an experimentally accessible spin asymmetry at future EIC and EicC facilities.
The $x$-dependent asymmetry exhibits a peak in the small-$x$ region, around $x\simeq 0.04$, and then decreases as $x$ becomes larger.

The sizable magnitude of the predicted asymmetry can be attributed to several factors. First, the model calculation of $\tilde D_T$ is proportional to the strong coupling $\alpha_s$. In the present work, both the model input and the subsequent QCDNUM evolution are initialized at the low model scale $\mu_0^2=0.23~\mathrm{GeV}^2$, where we take $\alpha_s(\mu_0^2)=0.817$. This value is larger than the effective couplings, $\alpha_s=0.2$ and $0.3$, adopted in previous spectator-model calculations~\cite{Mao:2014fma,Bacchetta:2007wc}, and consequently leads to a larger magnitude of $\tilde D_T$. This observation also indicates the model dependence of the predicted asymmetry associated with the choice of the effective coupling at the low input scale.
Second, the spin-dependent structure function contains the twist-3 kinematic factor $2M_\Lambda/Q$. Since the $\Lambda$ hyperon is relatively massive, with $M_\Lambda\simeq1.116~\mathrm{GeV}$, this factor can be numerically significant in the low- and moderate-$Q$ regions considered here. The relatively large magnitude of the predicted asymmetry is therefore not unexpected for $\Lambda$-hyperon production~\cite{Yang:2021zgy}.

We next examine the role of the strange-quark contribution.
In the present calculation, the flavor dependence of the numerator is governed by the product $f_1^q(x)\tilde D_T^{q\to\Lambda}(z)$. Consequently, the contribution from the $s\to\Lambda$ fragmentation channel is determined not only by the corresponding FF $\tilde D_T^{s\to\Lambda}$, but also by the strange-quark PDF $f_1^s(x)$ in the proton. As shown by the comparison between the solid and dashed curves in Figs.~\ref{fig:eic} and~\ref{fig:eicC}, including the strange-quark contribution reduces the magnitude of the negative asymmetry. This behavior is found for both EIC and EicC kinematics and can be traced to the opposite signs of the $u,d$- and $s$-quark contributions to $\tilde D_T^{q\to\Lambda}$. Although the strange-quark distribution in the proton is smaller than those of the valence $u$- and $d$-quarks, the strange channel nevertheless leads to a non-negligible modification of the predicted asymmetry. These results indicate that $A_{UUT}^{\sin\phi_{S_\Lambda}}$ may provide sensitivity to the strange-quark distribution $f_1^s(x)$ in the proton in future measurements at the EIC and EicC. A quantitative assessment of this sensitivity, however, would require a global analysis including the relevant experimental uncertainties and the theoretical uncertainties associated with the fragmentation functions.

\begin{figure}
  \centering
  \includegraphics[width=0.45\columnwidth]{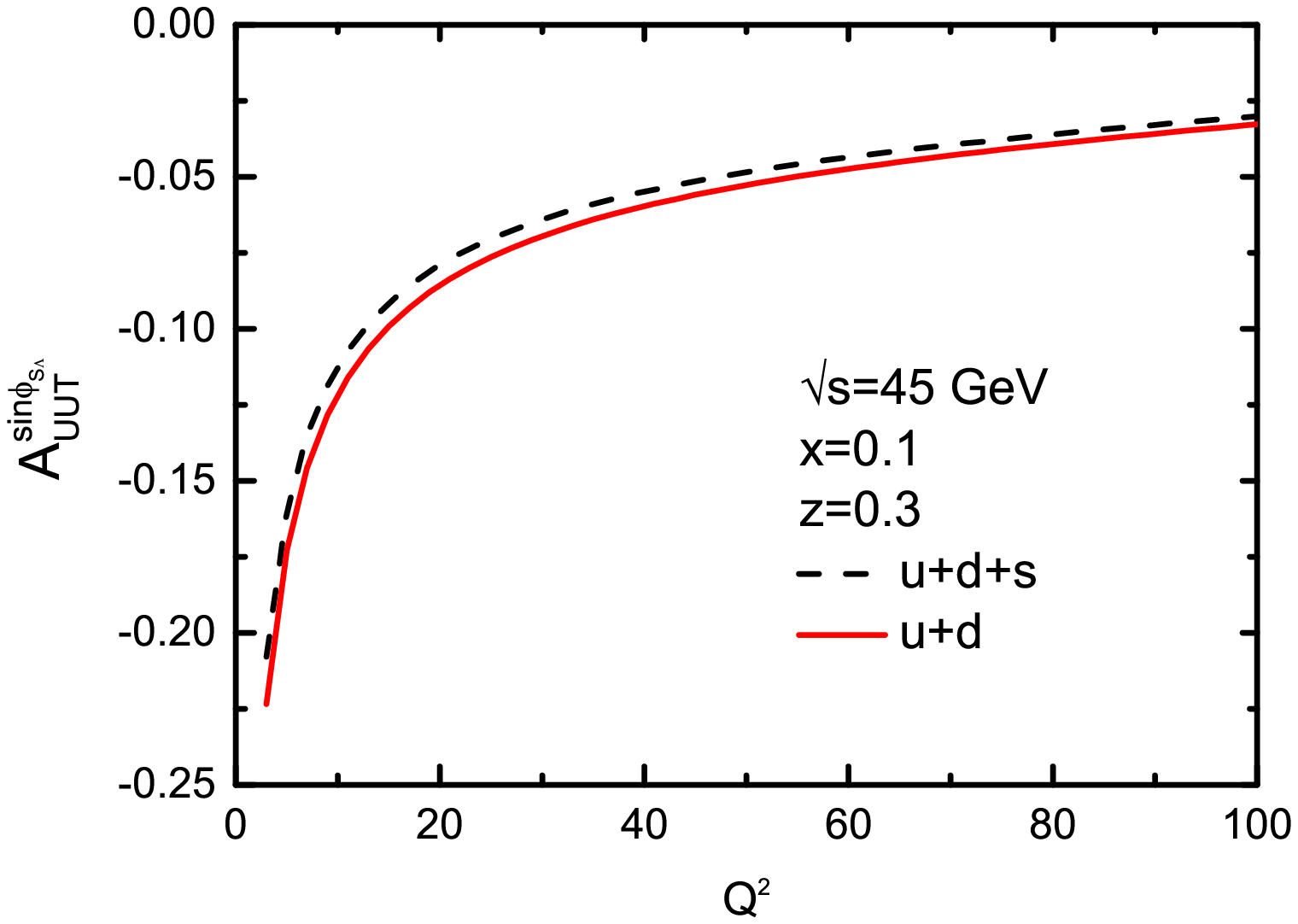}
  \includegraphics[width=0.45\columnwidth]{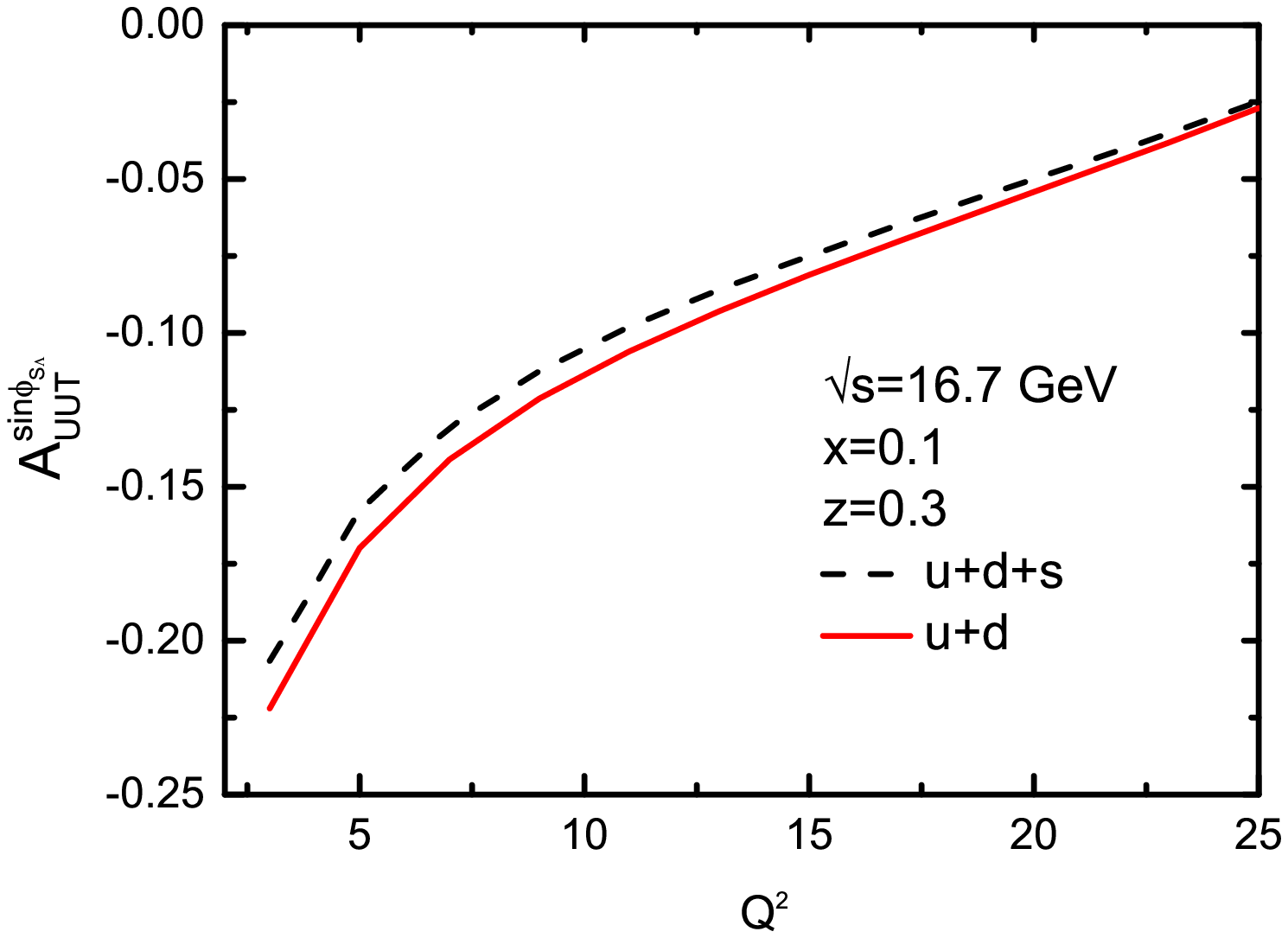}
  \caption{$Q^2$ dependence of the transverse SSA $A^{\sin\phi_{S_\Lambda}}_{UUT}$ including $u,d,s$ contributions (dashed lines) and $u,d$ (solid lines). The left and the right panels show the numerical results for EIC and EicC at $x=0.1, z=0.3$, respectively.}
\label{fig:q2}
\end{figure}

In Fig.~\ref{fig:q2}, we present the $Q^2$ dependence of the transverse-spin asymmetry $A^{\sin\phi_{S_\Lambda}}_{UUT}$ at fixed $x=0.1$ and $z=0.3$.
The predicted asymmetries are negative for both EIC and EicC kinematics, with their magnitudes decreasing as $Q^2$ increases. This behavior is primarily driven by the twist-3 kinematic factor $2M_\Lambda/Q$ appearing in the spin-dependent structure function.
The solid and dashed curves have the same meanings as in Figs.~\ref{fig:eic} and~\ref{fig:eicC}.
The comparison between the results with and without the strange-quark contribution shows that the strange channel also reduces the magnitude of the negative asymmetry in the $Q^2$-dependent results, consistent with the behavior observed in the $x$- and $z$-dependent asymmetries.
The choice $x=0.1$ provides a representative point in the moderate-$x$ region, where the valence $u$- and $d$-quark distributions remain dominant while the strange sea distribution is not negligible.
This choice therefore allows us to illustrate the relative impact of the strange-quark contribution on the $Q^2$ dependence of the asymmetry.
The value $z=0.3$ lies within the current fragmentation region and satisfies the kinematic cuts adopted for both the EIC and EicC.

For the EicC, the accessible $Q^2$ range is more restricted than that at the EIC because of its lower center-of-mass energy.
At fixed $x$, the DIS variables satisfy $Q^2=xys$.
With $\sqrt{s}=16.7~\mathrm{GeV}$, one has $s=(16.7~\mathrm{GeV})^2\simeq278.9~\mathrm{GeV}^2.$
Thus, at $x=0.1$, the upper limit on $Q^2$ imposed by the kinematic constraint $y<0.9$ is
$Q^2_{\rm max}=xys\simeq25.1~\mathrm{GeV}^2$.
This explains why the EicC result extends only to approximately $Q^2\simeq25~\mathrm{GeV}^2$. By contrast, the higher center-of-mass energy of the EIC allows access to a substantially broader $Q^2$ range at the same fixed value of $x$.

\section{Conclusion}\label{sec:Con}

In this work, we have investigated the single-spin $\sin\phi_{S_\Lambda}$ asymmetry for $\Lambda$-hyperon production in SIDIS off an unpolarized proton target.
After integrating over the transverse momentum of the final-state hadron, the asymmetry is described, within the collinear framework, by the product of the unpolarized PDF $f_1^q(x)$ and the twist-3, naive-T-odd qgq fragmentation function $\tilde D_T^{q\to\Lambda}(z)$. We have calculated the $\Lambda$-hyperon FF $\tilde D_T(z,\bm{k}_T^2)$ for different quark flavors within a diquark spectator model, with the model parameters determined from the corresponding fits. The resulting FFs exhibit a clear flavor dependence: $\tilde D_T^{u,d\to\Lambda}$ is negative, whereas $\tilde D_T^{s\to\Lambda}$ is positive and has a comparatively larger magnitude.

Using the resulting $\tilde D_T(z)$, we have estimated the asymmetry $A_{UUT}^{\sin\phi_{S_\Lambda}}$ for $\Lambda$ production at the kinematics relevant to the EIC and EicC.
Our numerical results indicate that the asymmetry is negative over the considered $x$, $z$, and $Q^2$ ranges and can attain a sizable magnitude.
We have also investigated the role of the strange-quark contribution. Although the strange-quark distribution in the proton is smaller than the valence $u$- and $d$-quark distributions, the strange channel produces a noticeable modification of the final asymmetry.
Thus, the inclusion of the strange-quark contribution reduces the magnitude of the negative asymmetry.
Our results suggest that the transverse-spin asymmetry $A_{UUT}^{\sin\phi_{S_\Lambda}}$ in $\Lambda$-hyperon production may be accessible in future measurements at the EIC and EicC. This observable provides a potentially useful probe of the twist-3 fragmentation function (FF) $\tilde D_T^{q\to\Lambda}$ and may also offer complementary sensitivity to the strange-quark parton distribution function (PDF) $f_1^s(x)$ in the proton within the collinear framework.

\section*{Acknowledgements}
This work was partially supported by the National Natural Science Foundation of China (NSFC) under Grants No. 12305088 and No. 12150013.

\end{document}